\documentclass[aps,prl,twocolumn,showpacs]{revtex4}
\usepackage{graphicx}

\begin{document}

\title{Temperature-induced reversal of magnetic interlayer exchange coupling}

\author{K.M. D\"obrich}
  \altaffiliation[present address: ] {Max-Born-Institut, D-12489 Berlin}
\author{M. Wietstruk}
  \altaffiliation[present address: ] {BESSY GmbH, D-12489 Berlin}
\author{J.E. Prieto}
 \email[Corresponding author. Email: ]{joseemilio.prieto@uam.es}
  \altaffiliation[present address: ] {Centro de Microan\'alisis de Materiales, 
Universidad Aut\'onoma de Madrid, E-28049 Madrid} 
  \author{F. Heigl}
  \author{O. Krupin}
  \author{K. Starke}\thanks{Deceased}
  \author{G. Kaindl}
  \affiliation{Institut f\"ur Experimentalphysik, Freie Universit\"at Berlin, Arnimallee 14, D-14195 Berlin, Germany \\
}

\date{\today}

\begin{abstract}

For epitaxial trilayers of the magnetic rare-earth metals Gd and Tb,
exchange coupled through a non-magnetic Y spacer layer, element-specific
hysteresis loops were recorded by the x-ray magneto-optical Kerr effect
at the rare-earth $M_5$ thresholds. This allowed us to quantitatively
determine the strength of interlayer exchange coupling (IEC). In addition
to the expected oscillatory behavior as a function of spacer-layer
thickness $d_Y$, a temperature-induced sign reversal of IEC was observed 
for constant $d_Y$, arising from magnetization-dependent electron 
reflectivities at the magnetic interfaces.

\end{abstract}


\maketitle

Thin magnetic layers coupled by magnetic exchange interaction across
a non-magnetic spacer layer have been studied extensively ever since
the first observations of antiferromagnetic (AFM) coupling in such
systems~\cite{majkr86,grun86,bai88,cebo89}. Widespread applications came through
the associated giant magnetoresistance effect (GMR)~\cite{bai88},
which made such layered magnetic structures essential elements in
advanced reading heads of magnetic storage devices. The interlayer
exchange coupling (IEC) behind these effects was found to oscillate
with the thickness of the non-magnetic spacer 
layer~\cite{pmr90,cebo291}.
This observation can be explained by the Rudermann-Kittel-Kasuya-Yosida (RKKY)
model, in which the oscillation period of the IEC is given by the length of
the extremal vectors that connect parallel sections of the Fermi
surface of the spacer layer~\cite{bruno91}. Dependences of the IEC on
other characteristic parameters of the magnetic layers, such as
thickness~\cite{bloemen94} and  composition~\cite{ebel98}, as well
as thickness of capping layers~\cite{deVries95} have also been
investigated.

In the past few years, several studies dealt with the temperature ($T$)
dependence of magnetic coupling through metallic as well as
insulating layers~\cite{lindner02,liuadenwalla03}. The standard RKKY
theory of IEC, considering constant magnetization of the magnetic
layers, leads to a rather weak $T$-dependence of the coupling~\cite{bruno95}. 
For metallic spacer layers,
thermal broadening of the Fermi edge causes a minor weakening of the
coupling strength. For insulating spacer layers, a similarly weak
increase in coupling strength is expected due to thermal population
of conduction-band states.
Thermal excitation of spin waves~\cite{almeida95} has also been
proposed to explain some of the experimental
observations~\cite{kalarickal07}. In a few cases, sizeable
$T$-dependences of IEC have been reported, e.g. for
Co/Cu/Co~\cite{persat97}, and unusual temperature behaviors have been
observed for ferromagnets exchange-coupled to antiferromagnets, like
NiO/Cu/NiFe~\cite{lin01}, or ferrimagnets, like
GdFe/Gd~\cite{demirtas04}. However, no $T$-induced sign reversal of 
IEC between ferromagnetic (FM) layers has been reported so far.

In this Letter, we report on a strong $T$-dependence of IEC
in epitaxial Gd/Y/Tb trilayers on W(110) that even leads to a
$T$-induced sign reversal of IEC for a given spacer-layer
thickness. This novel effect is explained by magnetization-dependent
electron reflectivities at the interfaces of Y with Gd and particularly 
with Tb. This causes strong temperature effects on the amplitude and on the 
{\em phase} of the oscillatory coupling strength as a function 
of spacer-layer thickness, leading to the observed sign reversals.

Heavy rare-earth metals are interesting magnetic materials due to
their different magnetic properties despite similar
crystalline and electronic structures. Gd and Tb (including Y)
crystallize in the hexagonal close-packed structure with lattice
parameters that differ by less than 2\%.
The spherical charge distribution of the 4$f^7$ shell of
Gd leads to a small magnetic anisotropy and hence to small
coercive fields in epitaxial films of good crystalline quality. On
the other hand, Tb has a large magnetic anisotropy due to its
aspherical charge distribution caused by a large atomic orbital
momentum ($L$=3). It is thus expected that the magnetization of
the softer Gd layer in the trilayer structures can be selectively
reversed.

\begin{figure}[t]
\includegraphics*[width=7.0cm]{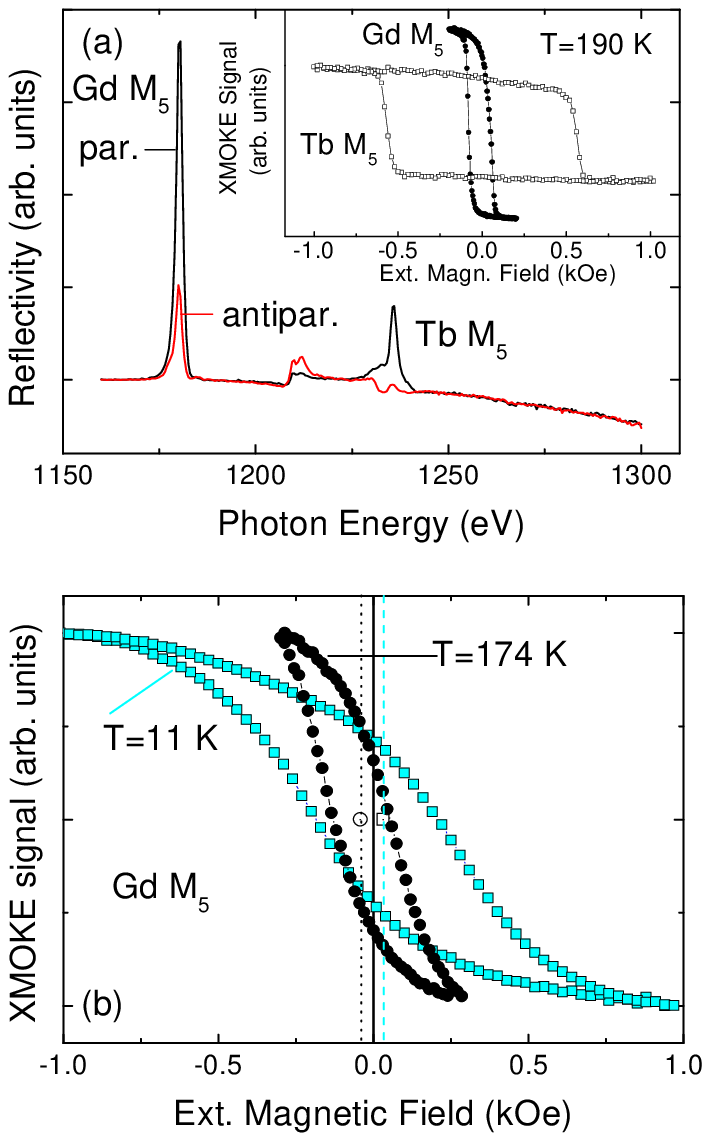}
\caption{\label{xmoke} (color online) (a) Soft x-ray reflectivity spectra 
in the 
region of the $M_{4,5}$ thresholds of Gd and Tb recorded from a Gd/Y/Tb/W(110)
trilayer with $d_Y$ = 1.5~nm at $T=$~20~K. The incidence direction of the 
circularly polarized x rays was nearly parallel and antiparallel to the 
in-plane sample magnetization.
The insert shows element-specific hysteresis curves
measured at photon energies corresponding to the Gd and Tb $M_5$
reflectivity maxima, respectively. (b) The Gd $M_5$ hysteresis curves,
measured at two different temperatures, are shifted due to exchange bias. 
For clarity, the centers of mass are marked with an open dot (at $T$ = 174~K) 
and an open square (at $T$ = 11~K),
showing negative and positive shifts, respectively.}
\end{figure}

Epitaxial Gd/Y/Tb/W(110) trilayers were grown in ultrahigh vacuum on a
W single-crystal substrate by metal-vapor deposition. Typical 
deposition rates were 1 to 4
monolayers (ML) per minute. The crystallinity of the layers was
checked by low-energy electron diffraction. The thicknesses of the
layers were $d_{Tb}$~=~10~nm, $d_{Gd}$~=~3.5~nm, with $d_Y$ ranging
from 0.3 to 3.3~nm. Trilayers with both constant $d_Y$ and a 
wedge-shaped Y~spacer layer were studied. The
as-grown films were annealed at temperatures known from previous
studies to result in smooth layers without significant
interdiffusion~\cite{heigl95}. Measurements of resonant soft
x-ray reflectivity with circularly polarized x-rays were performed
\textit{in situ} using the UE52 and UE56/1 beamlines at BESSY
(Berlin). The specularly reflected intensity was detected
by a Si photodiode mounted on a home-built $\theta-2\theta$ goniometer
inside the vacuum chamber. X-ray magneto-optical Kerr effect (XMOKE)
hysteresis loops were recorded by sweeping an external magnetic
field, produced by a rotatable magnet~\cite{magnet}, along the 
substrate $[110]$ direction that
corresponds to the $[1\bar{1}00]$ easy axis of magnetization of
epitaxial Tb/W(110) films~\cite{heigl95}.

The Gd/Y/Tb trilayers were cooled in an external magnetic field to
ensure saturation of the magnetically hard Tb layer. Fig.~\ref{xmoke}(a) 
shows typical reflectivity spectra of a remanently magnetized trilayer.
The large magnetic contrast allows to
perform XMOKE measurements in an element-specific way by
selecting the appropriate photon energy and by varying the applied
magnetic field. Typical hysteresis loops are displayed in the insert
of Fig.~\ref{xmoke}(a), reflecting the widely different coercivities of
Gd and Tb layers due to the large difference in anisotropies. 
This renders it possible to reverse only the magnetization
of the softer Gd layer by applying a magnetic field not strong
enough to influence the magnetization of the Tb layer. In these trilayers, 
the exchange coupling between the two magnetic layers acts as an 
effective bias field (exchange bias) that needs to be overcome in order
to reverse the magnetization of the softer layer. The resulting shift 
of the Gd hysteresis curve with respect to zero field is a measure of 
the coupling strength ~\cite{prieto05}. As an example,
Fig.~\ref{xmoke}(b) displays Gd hysteresis loops taken at two
different temperatures for $d_Y~=$ 1.5~nm: The data clearly show that 
the exchange bias changes sign from positive at 11 K to negative at 174 K.

\begin{figure}[b]
\includegraphics*[width=7.0cm]{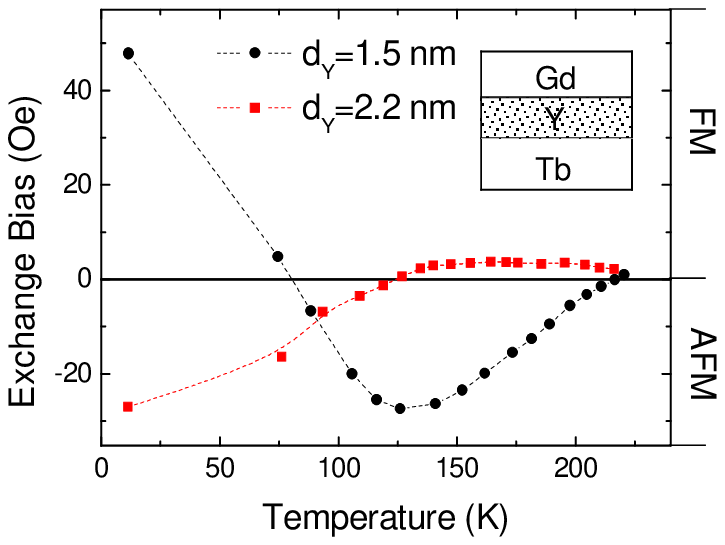}
\caption{\label{trilayer} (color online) Exchange bias as a function of 
temperature for Gd/Y/Tb/W(110) trilayers with two different $d_Y$. 
In both cases, the coupling changes sign with temperature.}
\end{figure}

\begin{figure}[t]
\includegraphics*[width=8.5cm]{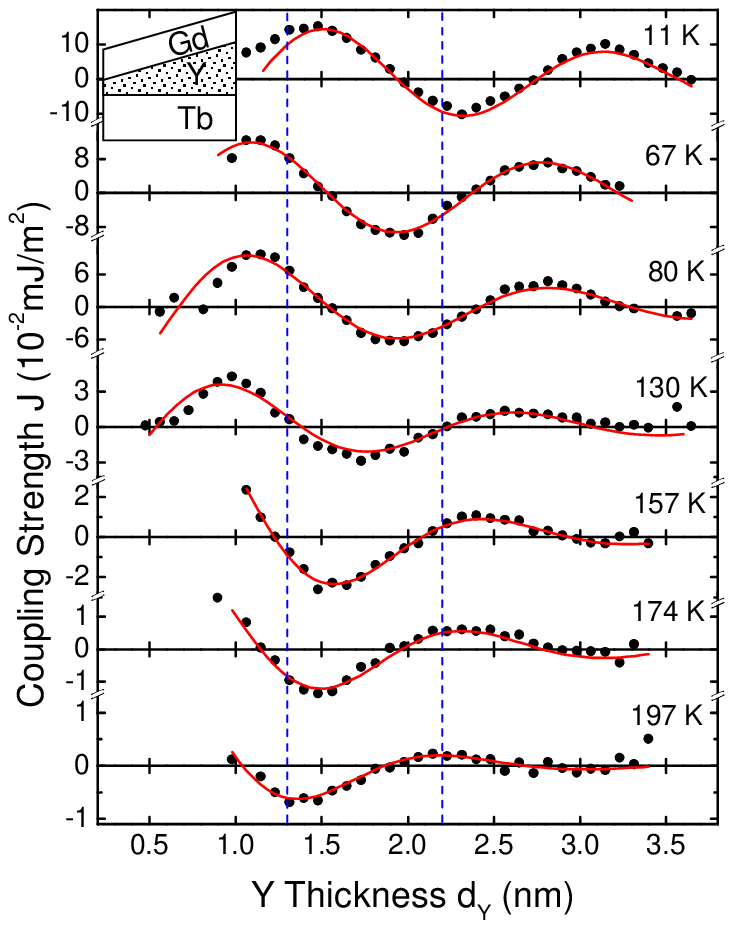}
\caption{\label{allcurves} (color online) Coupling strength $J$ between Gd 
and Tb layers for a Gd/Y/Tb/W(110) trilayer with a wedge-shaped Y-layer. 
Solid lines represent fits of eq.~(\ref{phenom}) to the data.
Note the scale changes on the ordinate for different $T$.}
\end{figure}

This striking $T$-dependence of the exchange bias was studied for 
11~K~$\leq T \leq~220$~K and for various $d_Y$. Fig.~\ref{trilayer} shows
the $T$-dependence of the exchange bias for two trilayers with fixed
$d_Y$: For $d_Y$=1.5~nm, the coupling changes from FM to AFM at $\approx$~80~K,
while for $d_Y$=2.2~nm, the opposite sign change from AFM to FM is observed 
at $\approx$~125~K. In both samples, the exchange bias vanishes at 
$\approx$~220~K.  

To investigate the sign change of the exchange bias more systematically,
i.e. as a function of $T$ and $d_Y$, trilayer structures with a 
wedge-shaped Y-spacer layer were studied. The results are summarized in
Fig.~\ref{allcurves}, where the coupling strength $J$ between the
magnetic Tb and Gd layers is displayed as a function of $d_Y$ for
various $T$. Here, $J$ is calculated as $J=H_b \, M_{Gd} \, d_{Gd}$, 
where the exchange bias $H_b$ was determined from the
shifts of the Gd hysteresis loops and $d_Y$ was varied by measuring at
different positions along the Y wedge (with an error $\Delta d_Y = 
\pm 0.1$~nm, due to the precision of $\Delta z~= \pm 0.1$~mm 
in the vertical position of the sample in the focussed x-ray beam).
An oscillatory behavior is evident for all temperatures, and the oscillation 
amplitudes decrease with increasing $d_Y$; in addition, the coupling 
strength decreases with increasing $T$, as expected for metallic spacer 
layers~\cite{bruno95}. 
The curves oscillate around $J$=0, demonstrating the absence of a significant
contribution of magnetostatic N\'eel coupling in our system~\cite{leal96}.

The novel result, evident from Fig.~\ref{allcurves}, is contained in
the significant, $T$-induced \textit{phase change} of the
oscillatory $J(d_Y)$ curves. This phase change causes the unusual
$T$-dependence of IEC, including the sign reversals with $T$ for 
constant $d_Y$ (see Fig.~\ref{trilayer}).
Further data for the $T$-dependence of the IEC for other values of
$d_Y$ can be obtained from Fig.~\ref{allcurves}.

The decay of the
amplitude is predicted by RKKY theory~\cite{bruno95} to follow a
$d_Y^{-2}$ law for large $d_Y$, in agreement with the present
results for $d_Y>1.5$~nm.
In order to describe the data also in the range $d_Y<1.5$~nm, we used an 
exponential dependence of the form $J\propto e^{-\beta d_Y}$.
A fit of the data in Fig.~\ref{allcurves} with the phenomenological 
expression~\cite{stiles93}
\begin{equation}\label{phenom}
J(d_Y) = A e^{-\beta d_Y} {\rm Im}(e^{2\pi i d_Y / \lambda } e^{-i\phi }),
\end{equation}
where the amplitude $A$, the decay constant $\beta$, the period 
$\lambda$, and the phase $\phi$ are taken as 
$T$-dependent adjustable parameters, results in  
the relevant parameters $\lambda$ and $\phi$.
The fit describes the experimental curves in Fig.~\ref{allcurves} rather
well, leading to an oscillation period of $\lambda~=~(1.5 \pm~0.1)$~nm, 
independent of $T$ and in good agreement with the
value of 1.6~nm determined from the length of the extremal vector in
[0001] direction of Y metal that connects parallel sections of the Fermi 
surface, as obtained in a recent band-structure calculation~\cite{gustav_unp}.

\begin{figure}[t]
\includegraphics*[width=7.5cm]{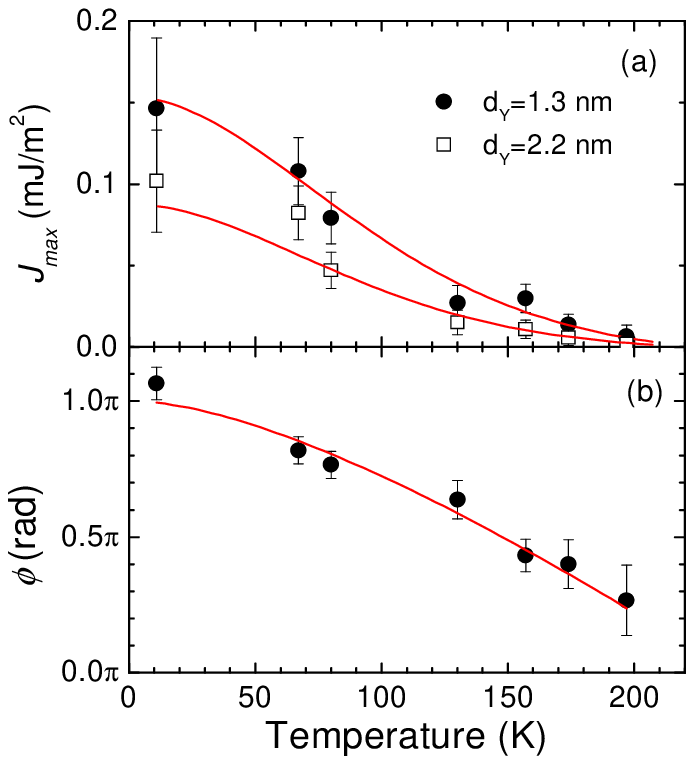}
\caption{\label{amplphase} (color online) $T$-dependence of (a) 
$J_{max}$ (for $d_Y =$ 1.3 and 2.2~nm) and (b) phase $\phi$ for Gd/Y/Tb 
trilayers, as extracted from data in Fig.~\ref{allcurves}.
For the fits, see text.}
\end{figure}

We now extract the $T$-dependences of the amplitudes and phases 
of the oscillations displayed in Fig.~\ref{allcurves}. As shown in 
Fig.~\ref{amplphase}, both the amplitude, $J_{max} = A \exp{(-\beta d_Y)}$, 
plotted for $d_Y = 1.3$ and 2.2~nm (see dashed vertical lines in 
Fig.~\ref{allcurves}) and the phase $\phi$ 
reveal strong changes with $T$ that cannot be explained by the 
standard RKKY theory~\cite{bruno95} or the spin-wave model~\cite{almeida95}. 
We therefore postulate that the
magnetizations of the layers have an intrinsic influence on the
coupling itself, as they are changing significantly in the studied
$T$ range, particularly for Tb metal (with a Curie temperature 
$T_C^{Tb}$=220~K as compared to $T_C^{Gd}$=293~K).

The origin of the indirect magnetic interlayer coupling through a
metallic layer can be understood within the picture of multiple
spin-dependent reflections of the valence electrons inside the
quantum well formed by the magnetic/nonmagnetic
interfaces~\cite{stiles93}. The difference $\Delta r$ of the complex
reflection coefficients for electrons of opposite spins causes a
polarization of the valence band of the spacer material that
mediates magnetic coupling. The phase accumulation model leads to an
expression for the coupling strength at $T~= 0$~K, $J(d_Y,0)$, which 
contains a factor $\Delta R_t e^{2i q_F d_Y}$~\cite{stiles93}, with $\Delta
R_t = |\Delta R_t| e^{i\phi} = \Delta r_{Tb} \Delta r_{Gd}$ (here, the
contribution of a single Fermi wave vector $q_F$ is considered). The
reflection coefficient at a magnetic interface depends on the
exchange splitting of the valence bands of the magnetic material.
Assuming a linear dependence in first-order approximation and
considering an exchange splitting proportional to the
saturation magnetization $M_S$, as given by the Stoner model, we
obtain
\begin{equation}\label{Deltar}
\Delta r(T) =  [ r^{\uparrow}(T) - r^{\downarrow}(T) ] \propto M_S(T).
\end{equation}
In this way, a $T$-dependence is introduced in the model
through the phases and amplitudes of the reflection coefficients.
Considering contributions from both magnetic layers, the coupling
strength $J$ will be approximately proportional to an effective
magnetization $M^*(T) = M_{Tb}(T) M_{Gd}(T) / [M_{Tb}(0) M_{Gd}(0)]$
[for $M_{Tb}(T)$ and $M_{Gd}(T)$, see Ref.~\onlinecite{rhy72}].
In addition, the $T$-dependence arising from thermal
broadening of the Fermi edge has to be taken into account, given
by $F(c\,T) = cT / \sinh(cT)$~\cite{bruno95}. 
Here, $c = a d_Y + b$, where
$a$ is a bulk term and $b$ an interface term. We obtain
$a$=$0.00018~({\rm \AA}{\rm K})^{-1}$ from the experimental oscillation 
period $\lambda$ (following Ref.~\onlinecite{bruno95}). The interface term $b$ 
is independent of $d_Y$~\cite{dAlbuquerque96,schwieger04} (see below).
Including the described temperature effects on the spin-dependent electron 
reflection coefficients, we obtain 
\begin{equation}\label{JpropM}
J(d_Y,T) \propto \exp{(2i q_F d_Y + i\phi)} F(cT) M^*(T).
\end{equation}
The amplitude of coupling, $J_{max}$, is well described by
this expression, as shown in Fig.~\ref{amplphase}(a).
A simultaneous fit of $J_{max}(T)=J_{max}(0)\,F(cT)M(T)$ to the 
data results in $J_{max}(0)$ = $(0.15 \pm 0.03)$ and 
$(0.09 \pm 0.03)$~${\rm mJ/m^2}$ for $d_Y =$ 1.3 and 2.2~nm, respectively, 
and $b= (0.015 \pm 0.002){\rm K}^{-1}$. These values are comparable to those
found for the Co/Ru/Co and Co/Cu/Co systems~\cite{schwieger04}.

The strong $T$-dependence of the IEC reported in this work is caused
by the $T$-dependent phase of the oscillations of $J(d_Y)$. It is
thus relevant to study this phase in more detail.
Figure~\ref{amplphase}(b) represents the $T$-dependence of the
phase  $\phi$ determined by least-squares fits of the
$J(d_Y, T)$ curves in Fig.~\ref{allcurves}. This strong $T$-dependence
of $\phi$ is caused by changes of the complex total
reflectivity $\Delta R_t$ with $T$. 
In first-order approximation and similar to the
case of $ |\Delta R_t| $, we assume that $\phi$ depends linearly on the
effective magnetization, $M^*(T)$. Accordingly, 
$\phi(T) = \phi_0 + \alpha M^*(T)$ was fitted to the data in  
Fig~\ref{amplphase}(b), leading to
$\phi_0 = (0.03\pm 0.08)\pi$ and $\alpha = (0.97 \pm 0.10)\pi$. The
relation $\phi(T) = \pi \cdot M^*(T)$ is thus compatible with the
experimental data within the error bars. It is tempting to interpret
this result analogously to the reflection of travelling waves from
free and fixed ends of a vibrating rope, with phase changes of
0~and~$\pi$, respectively. The vanishing magnetic exchange splitting
of the valence bands at $T_C$ is intuitively expected to result in a
reflection of the electron waves with zero phase change. While the
value of $\phi$ close to $\pi$ at $T~= 0$~K is possibly accidental, 
its vanishing at the Curie temperature of Tb strongly
supports our assumption that the variations in magnetizations of the
magnetic layers cause the observed strong $T$-dependence of IEC in 
the present case.

In summary, the observed $T$-induced sign reversal of
magnetic interlayer exchange coupling is described by
the thermal variations of the magnetizations of the two
magnetic layers, which lead to strong effects on amplitudes and
phases of the spin-dependent electron reflectivities at the
interfaces. This new effect might find practical applications in
temperature-sensitive devices through the GMR associated with a
reversal of IEC. The working temperature range can be tuned by selecting 
suitable materials and layer thicknesses.

Work supported by the DFG project KA 564/10-1. The authors
thank H.-Ch. Mertins, K. Godehusen and J. N\"otzold for experimental 
support. J.E.P. acknowledges financial support by the A.-v.-Humboldt
foundation and by the MEC program ``Ra\-m\'on y Cajal''. 

\bibliography{gdytb}

\end{document}